\documentclass[%
 reprint,
 amsmath,amssymb,
 aps,
prd,
]{revtex4-2}

\usepackage{graphicx}% Include figure files
\usepackage{dcolumn}% Align table columns on decimal point
\usepackage{bm}% bold math
\newcommand{\Msun}{${\rm M}_{\odot}$\ }

\newcommand{\mnras}{Monthly Notices of the Royal Astronomical Society}

\newcommand{\aap}{Astronomy and Astrophysics}
\newcommand{\araa}{Annual Review of Astronomy and Astrophysics}
\newcommand{\jcap}{Journal of Cosmology and Astroparticle Physics}

\begin{document}

\preprint{APS/123-QED}

%\title{On the "Search for lensing signatures in the gravitational-wave observations from the first half of LIGO-Virgo’s third observing run"}% Force line breaks with \\
\title{Evidence for lensing of gravitational waves from LIGO-Virgo}% Force line breaks with \\
%\thanks{A footnote to the article title}%

\author{J.M. Diego}
 %\homepage{http://www.cosmicspectator.org}
 \email{jdiego@ifca.unican.es}
  %\altaffiliation[Also at ]{Physics Department, XYZ University.}%Lines break automatically or can be forced with \\
%\author{Second Author}%
% \email{Second.Author@institution.edu}
\affiliation{
Instituto de F\'isica de Cantabria (CSIC-UC) Edificio Juan Jord\'a. Avda Los Castros s/n. 39005 Santander, Spain.}
\author{T. Broadhurst}
\affiliation{
Department of Theoretical Physics, University of the Basque Country UPV-EHU, 48040 Bilbao, Spain. \\
Donostia International Physics Center (DIPC), 20018 Donostia, The Basque Country, Spain. \\
IKERBASQUE, Basque Foundation for Science, Alameda Urquijo, 36-5 48008 Bilbao, Spain.
}
\author{G. F. Smoot}
\affiliation{
IAS TT \& WF Chao Foundation Professor, IAS, Hong Kong University of Science and Technology, Clear Water Bay, Kowloon, 999077 Hong Kong, China. \\
Paris Centre for Cosmological Physics, Universit\'{e} de Paris, emertius, CNRS,  Astroparticule et Cosmologie, F-75013 Paris, France A, 10 rue Alice Domon et Leonie Duquet,75205 Paris CEDEX 13, France. \\
Donostia International Physics Center (DIPC), 20018 Donostia, The Basque Country, Spain. \\
Physics Department and Lawrence Berkeley National Laboratory, University of California, emeritus Berkeley,94720 CA, USA. 
 %\textbackslash\textbackslash
}%

%\collaboration{MUSO Collaboration}%\noaffiliation

%\author{Charlie Author}
% \homepage{http://www.Second.institution.edu/~Charlie.Author}
%\affiliation{
% Second institution and/or address\\
% This line break forced% with \\
%}%
%\affiliation{
% Third institution, the second for Charlie Author
%}%
%\author{Delta Author}
%\affiliation{%
% Authors' institution and/or address\\
% This line break forced with \textbackslash\textbackslash
%}%

%\collaboration{CLEO Collaboration}%\noaffiliation

\date{\today}% It is always \today, today,
             %  but any date may be explicitly specified

\begin{abstract}
%Recently, the LIGO-Virgo collaboration published a work where they conclude that there is no evidence for lensed gravitational waves in the first half of the O3 run. We discuss possible issues with this (and earlier) conclusion(s), related with their choice of priors, and argue that instead, LIGO-Virgo Collaboration's analysis does indeed show evidence for pairs of lensed gravitational waves. Further evidence will be available soon with the publication of the new data from the second half of the O3 run.  

%{\bf Alternative ABSTRACT}
Recently, the LIGO-Virgo Collaboration (LVC) has concluded there is no evidence for lensed gravitational waves (GW) in the first half of the O3 run \citep{LIGO2021}, claiming ``We find the observation of lensed events to be unlikely, with the fractional rate at $\mu > 2$ being $3.3 \times 10^{-4}$”. While we agree that the chance of an individual GW event being lensed at $\mu > 2$ is smaller than $<10^{-3}$, the number of observed events depends on the product of this small probability times the rate of mergers at high redshift.
Observational constraints from the stochastic GW background indicate that the rate of conventional mass BBH mergers ($8 < M /{\rm M}_{\odot} < 15$) in the redshift range $1 < z < 2$  could be as high as O($10^7$) events per year, more than sufficient to compensate for the intrinsically low probability of lensing. 
To reach the LVC trigger threshold these events require high magnification, but would still produce up to 10 to 30 LVC observable events per year. Thus, all the LVC observed ordinary stellar mass BBH mergers from this epoch must be strongly lensed with as many as $10^6\,yr^{-1}$ events lying below the current detection threshold. By adopting a low BBH coalescence rate at high redshift, LVC implicitly assume that lensed events are insignificant and thus incorrectly underestimate the distances of most BBH events and correspondingly overestimate masses by factors of 2 to 5.
Furthermore, the LVC adopted priors on time delay for ideal circularly symmetric lenses 
are in tension with the known distribution of observed time delays of lensed quasars that require elliptical potentials
with a broad spread of time delays. Pairs of events like  GW190421\_213856–GW190910\_112807 and GW190424\_180648–GW190910\_112807, 
which are directly assigned a probability of zero by LVC, should instead be considered as candidate lensed BBH pairs, since their 
separation in time is typical of lensed quasars. Replacing the LVC model prior for the time delay distribution with the empirical 
Quasar-based distribution reverses the LVC conclusions and says that a significant fraction of BBH pairs identified by LVC are 
viable multiply-lensed events, including quadruple systems.

\end{abstract}

%\keywords{Suggested keywords}%Use showkeys class option if keyword
                              %display desired
\maketitle

%\tableofcontents

\section{\label{sec_Intro}Introduction}
%%%%%%%%%%%%%%%%%%%%%%%%%%%%%%%%%%%%%%%%

In a recent paper, \cite{LIGO2021}, the LIGO-Virgo collaboration (LVC) consider the possibility that some of the detected gravitational waves are magnified by gravitational lensing. If lensing is involved, the strain of a distant gravitational wave (or GW) would be amplified by $\sqrt{\mu}$, where $\mu$ is the amplification factor. In this case, and if lensing is ignored, the inferred luminosity distance is smaller by the same factor $\sqrt{\mu}$. Since LVC measures redshifted masses, the error in the luminosity distance (and consequently on the inferred redshift) may significantly overestimate the intrinsic masses by a factor $(1+z_t)/(1+z_i)$, where $z_t$ is the true redshift of the binary emitting the gravitational waves, and $z_i$ is the incorrectly inferred redshift. For typical lensing configurations found in astrophysical observations, this factor is usually between 2 and 5, as shown by  \cite{Broadhurst2018,Broadhurst2019,Broadhurst2020a,Broadhurst2020b} (referred hereafter as the BDS model and that we discuss in more detail in section \ref{sec_BDS}). 

The LIGO-Virgo team mentions that, ``We find the observation of lensed events to be unlikely, with the fractional rate at $\mu >2$ being $3.3 \times 10^{-4}$". 
We agree with this statement, but note that a factor $\approx 3$ uncertainty in the precise value of this fractional rate is expected, due to variability in the density profile of the lenses, concentration, and substructure. Nevertheless, if the rate of mergers at high redshift is larger than $10^4$ Gpc$^{-3}$ yr$^{-1}$, observation of lensed GWs by LIGO-Virgo would not only be possible, but inevitable. The role of lensing in the current early stages of GW astronomy can be considered similar to the early days of the far infrared (IR) sub-mm astronomy with Herschel, where the brightest sub-mm galaxies were all found to be strongly lensed \cite{Negrello2010}. This is typical in gravitational lensing of background populations with a steep luminosity function. In the case of GWs, the role of sub-mm luminosity is played by the chirp mass, while the role played by the negative k-correction in sub-mm galaxies can be attributed, in the case of GWs, to a rapidly evolving (with redshift) merger rate between z=0 and z=2. In flux limited surveys with relatively high thresholds (like LVC), the combination of a high merger rate at $z>1$ (where the optical depth of lensing is greater) with large magnification factors result in a significant number of strongly lensed GWs. \\ 

A basic prediction from strong lensing is that at magnification factors larger than $\approx 5$ multiple images are always produced. For a GW detector, these multiple events would appear as GWs from the same sky location, and with similar strains, \citep[except for a possible small phase shift between images with different parity, as shown by][] {Dai2017}. Due to a possible difference in the magnification factors, the observed events may appear as having different SNR, and hence could be interpreted as having different intrinsic redshift and masses (but similar redshifted mass). In some cases, some of the multiply lensed images may not be observed due to unfavourable beam orientations of the LIGO-Virgo detectors, and/or a smaller magnification for one of the images.
Based on the similarity of the posteriors and the sky location, LVC defines a factor $\mathcal{B}^{\rm overlap}$ that measures the similarity between GWs , by evaluating the overlap of posterior distributions from  the individual events, including the coincidence in sky location. In \cite{LIGO2021}, several candidates are identified having a large  $\mathcal{B}^{\rm overlap}>50$, but these are all later rejected by their separation in arrival time, or time delay due to gravitational lensing. This probability, or penalty function, is parameterized by the quantity  $\mathcal{R}^{\rm gal}$. In this paper we argue that while the calculation of $\mathcal{B}^{\rm overlap}$ is robust, the same can not be said about  $\mathcal{R}^{\rm gal}$, where we identify several issues that bias the conclusions in \cite{LIGO2021}. In a more refined analysis, the LVC arrive to similar conclusions, but these too are affected by similar biases in their choice of priors. 

In \cite{LIGO2021}, LVC searches for lensed pairs of GWs in the 39 candidate events reported in the GWTC-2 catalog from the first half of the O3 run \cite{LIGO2020GWTC2}. They find that 19 out of the original 39 candidates (48.7\%)  (or 2.6\% if one considers all possible 740 combinations of the 39 events) can form viable lensed candidates pairs, based solely on the similarity between parameter posteriors and sky localization. This large fraction of candidates having high-consistency between them suggest some level of correlation between the GWs. It is natural to consider lensing as a possible explanation for this correlation, motivating the detailed search made by LVC.  In order to test this hypothesis, the LIGO-Virgo collaboration performs a series of analyses to asses the likelihood that some of the candidates from the GWTC-2 catalog are being multiply lensed. Below we discuss their results and main conclusions. 
 
%In particular, they identify 19 lensed pair candidates having high similarity in space parameter and consistency in sky localization. Among these, in 5 pairs each one of the GW in the pair has a 90\% credible region in the sky, $\Delta \Omega \leq 2000$ deg$^2$, or less than 5\% of the sky. 
%This large fraction of lensed candidates with similar parameters and sky location goes against the naive expectation\footnote{if lensing rates are low, as predicted by many models except for the BDS model that has predicted that the fraction of lensed events is $\approx 50\%$ the total of observed events} that only a few of the 39 candidates in the GWTC-2 catalog would end up been viable pairs. On the other hand, this high rate agrees with the one predicted by the BDS model under the lensing hypothesis. 

\section{Stochastic background}
%%%%%%%%%%%%%%%%%%%%%%%%%%%%%%%%
The first analysis can be found in section 3.3 of \cite{LIGO2021}, were they present a discussion based on the non-detection of the stochastic background of GWs as evidence that the merger rate at high redshift is too low to produce significant lensed events. Here we briefly comment that this conclusion is derived from a model that assumes as prior information that lensing is not taking place. \cite{LIGO2021} follows the work of \cite{Buscicchio20}, but in that reference (as well as in \cite{LIGO2021}) the model that is being constrained is normalized to the observed local rate. This implicitly assumes that none of the observed events are lensed, but also that the contribution of the local universe $(z<0.3)$ to the stochastic background is significant. A model that assumes a much smaller rate in the local universe, but a much larger rate at high redshift, can simultaneously produce lensing events and keep the stochastic background of GWs relatively low.
Since the stochastic background of GWs has a strong contribution from local events and massive binaries, reducing the rate of local events and lowering the mass of the binaries, allows an increase in the rate of distant events, while maintaining this background below current limits. Luckily,  a model that allows lensing to happen was checked against the stochastic background of GWs in \cite{Mukherjee2021}, and found to be consistent with the current limit of the stochastic background of GWs. This work is cited by \cite{LIGO2021}, but surprisingly without comment about the compatibility of this model with the current upper limit from the stochastic background. Models that allow for signifcant events in LIGO-Virgo being strongly lensed are actually close to the current limit (but still below). Hence observations in the near future of the stochastic background will soon be able to confirm or rule out this type of model.

%The main difference between the BDS model and the models considered by the LIGO-Virgo collaboration is the binary merger rate as a function of redshift. Its main characteristic is the high rate of mergers between z=1 and z=2, and the relatively low rate in the local universe. This low local rate allows the BDS model to be consistent with constraints derived from the stochastic background of gravitational waves as shown in \cite{Mukherjee2021}. Future observations will soon be able to reject or confirm this model (if the stochastic background is detected at the level predicted by the BDS model). 

\section{Posterior overlap analysis}
%%%%%%%%%%%%%%%%%%%%%%%%%%%%%%%%%%%%%%%
The second analysis is presented in section 5.1 of \cite{LIGO2021}, where the LIGO-Virgo collaboration performs a posterior-overlap analysis of the 39 candidate events reported in the GWTC-2 catalog. The score in  \cite{LIGO2021} used to classify pairs of events as candidates to multiply imaged events, is given by the product of two independent functions , $\mathcal{B}^{\rm overlap}$ and $\mathcal{R}^{\rm gal}$, where $\mathcal{B}^{\rm overlap}$ quantifies the similarity between the posteriors of the analysed GWs and coincidence in sky location (a basic prediction from lensing), while $\mathcal{R}^{\rm gal}$ measures the consistency between the observed separation in arrival times and the expected one from lensing. Since time delays and magnification are strongly correlated, this correlation would introduce also a correlation in the inferred distances. The LIGO-Virgo team avoids this possible correlation between  $\mathcal{B}^{\rm overlap}$ and $\mathcal{R}^{\rm gal}$ by not including the posteriors of the inferred distances in the overlap analysis. This is particularly fortunate for our purposes, since most of the uncertainties in lens modelling can be attributed to the term $\mathcal{R}^{\rm gal}$.  LVC find 19 pairs of events (among the 740 possible combinations from the original 39 events) with a large value of $\mathcal{B}^{\rm overlap}$, but they are all later heavily penalized by a small value $\mathcal{R}^{\rm gal}$, which depends on the time separation between the two GWs in the pair.  More precisely, $\mathcal{R}^{\rm gal}$ is defined as follows;
\begin{equation}
    \mathcal{R}^{\rm gal} = \frac{p(\Delta t|\mathcal{H}_{SL})}{p(\Delta t|\mathcal{H}_{U})}
    \label{Eq1}
\end{equation}
where $p(\Delta t|\mathcal{H}_{SL})$ and $p(\Delta t|\mathcal{H}_{U})$  are the prior probabilities of the time delay under the strongly lensed and unlensed hypotheses, respectively. Few details are given about $p(\Delta t|\mathcal{H}_{U})$ (and not enough about $p(\Delta t|\mathcal{H}_{SL})$). It is just mentioned that  $p(\Delta t|\mathcal{H}_{U})$ is derived assuming GWs follow a Poissonian process.  More information can be found in \cite{Haris2018} which serves as the inspirational work for the definition of $\mathcal{R}^{\rm gal}$. In that work a distribution is shown for $p(\Delta t|\mathcal{H}_{U})$ in their Figure 2, that peaks in $\approx 40$ days. In the same figure the distribution for  $p(\Delta t|\mathcal{H}_{SL})$ peaks in $\approx$ 1 hour. This is in tension with the known distribution of time delays from QSOs as discussed below \citep[see also][]{Millon2020,Millon2020B}. Although it is speculative on our side to assume that similar distributions are adopted in \cite{LIGO2021}, the continuous reference to \cite{Haris2018}, and the conclusions reached by the LIGO-Virgo collaboration make us suspect that the priors $p(\Delta t|\mathcal{H}_{SL})$ and $p(\Delta t|\mathcal{H}_{U})$ do not differ much from those in \cite{Haris2018}. 

\begin{figure}[ht]
\includegraphics[width=9cm]{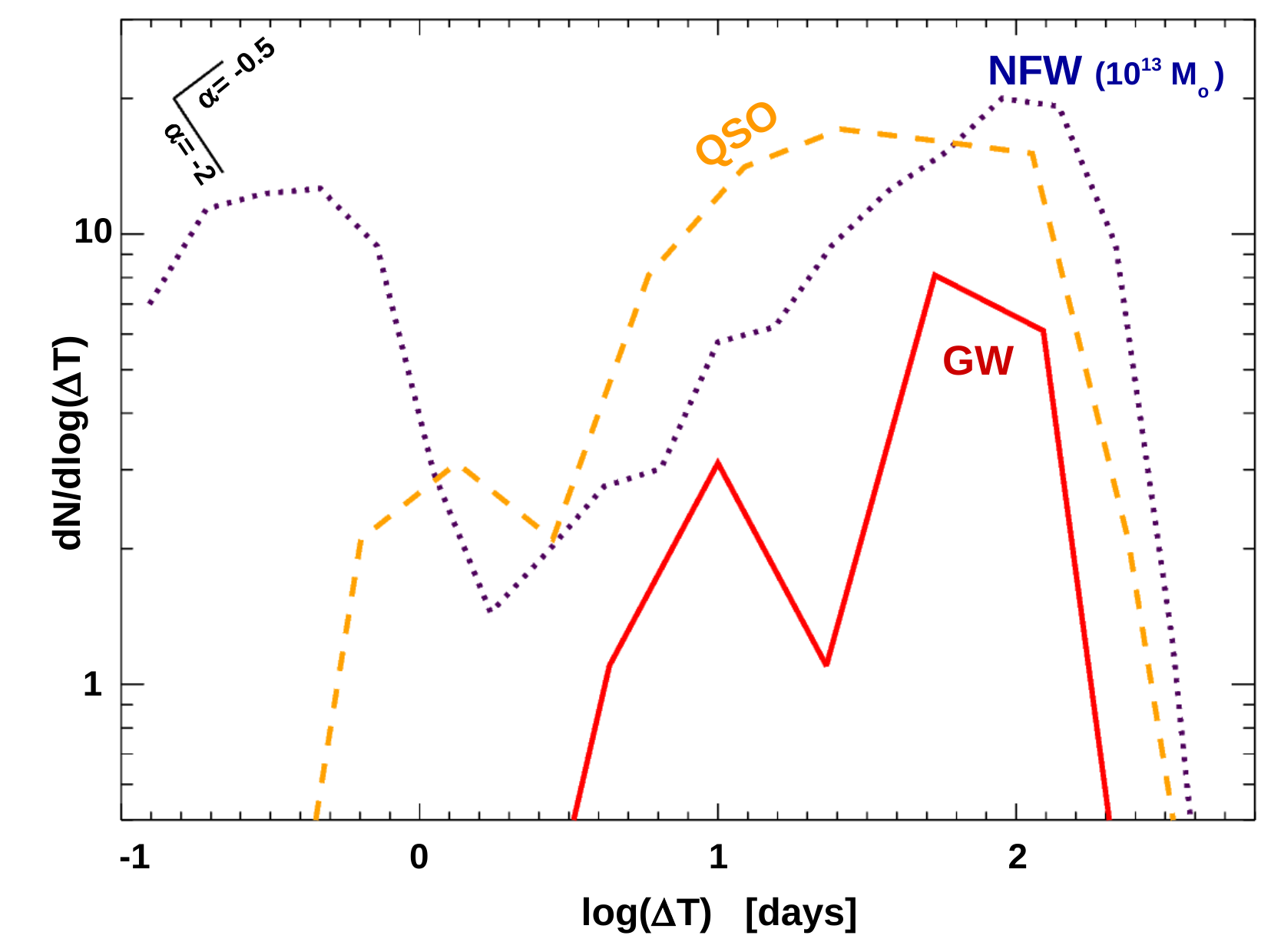} 
\caption{\label{fig1} Observed distribution of time delays of lensed QSOs (orange dashed line) from \cite{Millon2020,Millon2020B}, compared with the prediction from one NFW lens with ellipticity e=0.3,  mass $10^{13}$ M$_{\odot}$, located at $z_l=0.4$ and for a background source at $z_s=2$ (dark blue dotted line).
%, and a similar lens but with mass $3\times 10^{12}$ M$_{\odot}$ (light blue dotted line). 
Only pairs of lensed images where the minimum magnification in one of the images is larger than 5 are considered. 
%Time delays for QSOs smaller than a few days are difficult to measure due to the uncertainty in the time delay measurement. 
The red solid line shows the distribution of time intervals for the LVC lensed candidate pairs in \cite{LIGO2021}. This distribution peaks at time delays  similar to the case of the observed lensed QSOs. The two blue dots in the bottom right part of the plot show the time delay of the two pairs of events which were directly excluded by LVC on the (erroneous) basis that they are not consistent with being lensed owing to their large time delay.}
\end{figure}

To compute time delay statistics,  \cite{LIGO2021} considers the Singular Isothermal Sphere (or SIS) model (and also to compute rates, but rates are less relevant for this discussion). This is in general a good choice for generic lens modelling (except in the low-end and high-end of the mass spectrum where the NFW is a better description). In \cite{LIGO2021}, only masses in the galaxy scale are being considered, ignoring the contribution from more massive halos like groups and clusters, which naturally predict longer time delays (typically years as opposed to days or months for galaxies). Neglecting the contribution from clusters is made on the basis that their contribution to the optical depth is low. However, this is only true for the most massive clusters. Less massive halos, even though they are less numerous than regular galaxies, contribute individually more to the lensing optical depth owing to their larger mass as shown by \cite{Hilbert2007,Diego2019}, where halos with mass a few times $10^{13}$ M$_{\odot}$ at redshift $z \approx 0.5$ make a significant contribution to the lensing optical depth. As a demonstration, we show a typical  $10^{13}$ M$_{\odot}$ lens model in Figure \ref{fig1} (dotted line), which can be taken as a representative lens for the observed distribution of time delays in quasars (or QSO, dashed line). The SIS model has also (by construction) zero ellipticity. This assumption has consequences for the derived time delays. Real lenses found in nature have ellipticites and/or are subject to external shears. Ellipticicty plays an important role in lensing time delays, since elliptical lenses can produce more than two images, and broadens the distribution of time delays relative to SIS. In lenses with no ellipticity (like the SIS model), the lens forms a point-like caustic, and only two images are produced (a central third image is demagnified and unobservble. When ellipticity (or external shear) is introduced, the point-like caustic becomes a diamond-shape with four cusps. The size of the diamond-shape region grows with ellipticity and/or external shear and with lens mass. The size of the radial critical curves depend also on the mass and concentration of the halo. If the source is placed inside the diamond-shape region, two more images form, increasing the number of time delays between multiple images. Time delays between the two newly formed images are usually short when the source is near the caustic, and longer when the source is near the centre of the caustic region. Time delays between the tangential and radial images are usually significantly longer, but in this case, tangential images have usually modest magnification factors and are more difficult to be observed.   

%The LIGO-Virgo team does not provide sufficient information regarding the distribution used for the time delay, so it is difficult to precisely asses possible issues with the assumed distribution of time delays. Continuous reference is made to the work of \cite{Haris2018}, which sets the base for the analysis presented by the LIGO-Virgo team. In \cite{Haris2018}, their figure 2 shows the PDF of time delays peaking at $\approx 1$ hour. 
Figure 3 in \cite{LIGO2021} offers clues regarding the time delay distribution adopted by the LIGO-Virgo collaboration. By comparing pairs of events with similar $\mathcal{B}^{\rm overlap}$, the figure shows how pairs of GWs with temporal separations of $\approx 2$ months are heavily penalized (i.e smaller values $\mathcal{R}^{\rm gal}$) compared with pairs of GWs with temporal separations of days. This too goes against our prior information of time delay statistics, where approximately half the known time delays are above 1 month. Also, theoretical predictions suggest that time delays of months are expected to be common \citep{Oguri2002,Kuhlen2004,Li2012}. Simple predictions can be derived for the SIS model adopted by LVC. 
For this model it is well known that the time delay between the two images is simply
\begin{equation}
    \Delta T_{SIS} = \frac{(1+z_l)}{2c}\frac{D_lD_s}{D_{ls}}(\theta_1^2 - \theta_2^2) 
\end{equation}
where $D_l$, $D_s$, and $D_{ls}$ are the angular diameter distances to the lens, to the source, and from the lens to the source respectively; $z_l$ is the redshift of the lens; and $\theta_1$, $\theta_2$ are the angular distances from each image to the centre of the lens. 
Typical lenses found in nature have Einstein radii $\Theta_E \approx 1-2$ arcsecond. Meanwhile $\theta_1$, $\theta_2$ are usually similar to $\Theta_E$ with a difference ${\rm abs}(|\theta_1| - |\theta_2|) \approx 0.5$ arcsec \citep{Oguri2003}, resulting in time delays of approximately one week for these galaxies. 
More massive lenses, although less numerous, have a larger cross section for lensing (i.e they can magnify a larger area in the source plane), so they can compensate their smaller number with their larger cross section, resulting in a significant number of time delays between months (small groups) to years (clusters). The time delay for these heavier lenses scales also with the mass of the lens. 

%
% Made by LIGO/Plot_TimeDelay.pro in 2015 Toshiba computer

Instead of a prior given by the simple SIS model, a data-driven prior (following the spirit of a truly Bayesian framework) could be taken directly from  observations of known lenses. In Figure \ref{fig1} we show the observed distribution of QSO time delays (dashed line) as compiled by \cite{Millon2020}. Taking logarithmic bins in time delay, the probability peaks at $\approx 3$ months. For illustration purposes,  we show also the predicted time delay computed numerically for an elliptical (e=0.3) NFW halo with mass $10^{13}$ M$_{\odot}$, at $z_l=0.4$ and for a background source at $z_s=2$ (dark blue dotted line). Since time delays depend on the magnification, for this lens model we consider only the brightest pair of images, and with the added constrain that the minimum magnification of one of the two images is at least 5. As discussed earlier, this type of halo and redshift is expected to contribute significantly to the lensing optical depth for background sources at $z_s=2$.  
 For this model, an additional small peak (not shown) exists at time delays of few years corresponding to differences between the brightest radial image and the third image outside the Einstein ring. However, these images have relatively low magnifications, so they are less likely to be detected in real flux-limited observations. 
 %A smaller analytical halo with a mass $\approx 3$ times smaller is also shown (light blue dotted line). Lighter halos predict smaller time delays, but time delays of months are still possible. For this lighter halo, the small peak corresponding to the radial-tangential pairs (with $\mu_{min}>5$) can be appreciated at $\approx 140$ days.  
 Finally, the red solid line shows the distribution of time separations between the 19 lensed pair candidates in \cite{LIGO2021}. Here we emphasize that the distribution of observed QSO time delay (orange dashed line) and the distribution of time intervals between lensed GW pairs (red solid line) are similar. Should the prior $p(\Delta t|\mathcal{H}_{SL})$ be taken as the orange dashed line, and the prior $p(\Delta t|\mathcal{H}_{U})$ as the red solid line (under the hypothesis that these are the real time intervals between valid unlensed candidates), the ratio $\mathcal{R}^{\rm gal}$ would be approximately constant, playing a much less relevant role in penalizing pairs of lensed candidates.  
 The two small solid lines in the top-left corner show the slope, $\alpha$, of probabilities scaling as $\Delta T^{\alpha}$. The case $\alpha=-0.5$ agrees well with the observed QSO time delay distribution and theoretical expectation for the NFW model. The case $\alpha=-2$ is the inferred distribution of $\mathcal{R}^{\rm gal}$  as discussed below. 

\begin{figure}[ht]
\includegraphics[width=9cm]{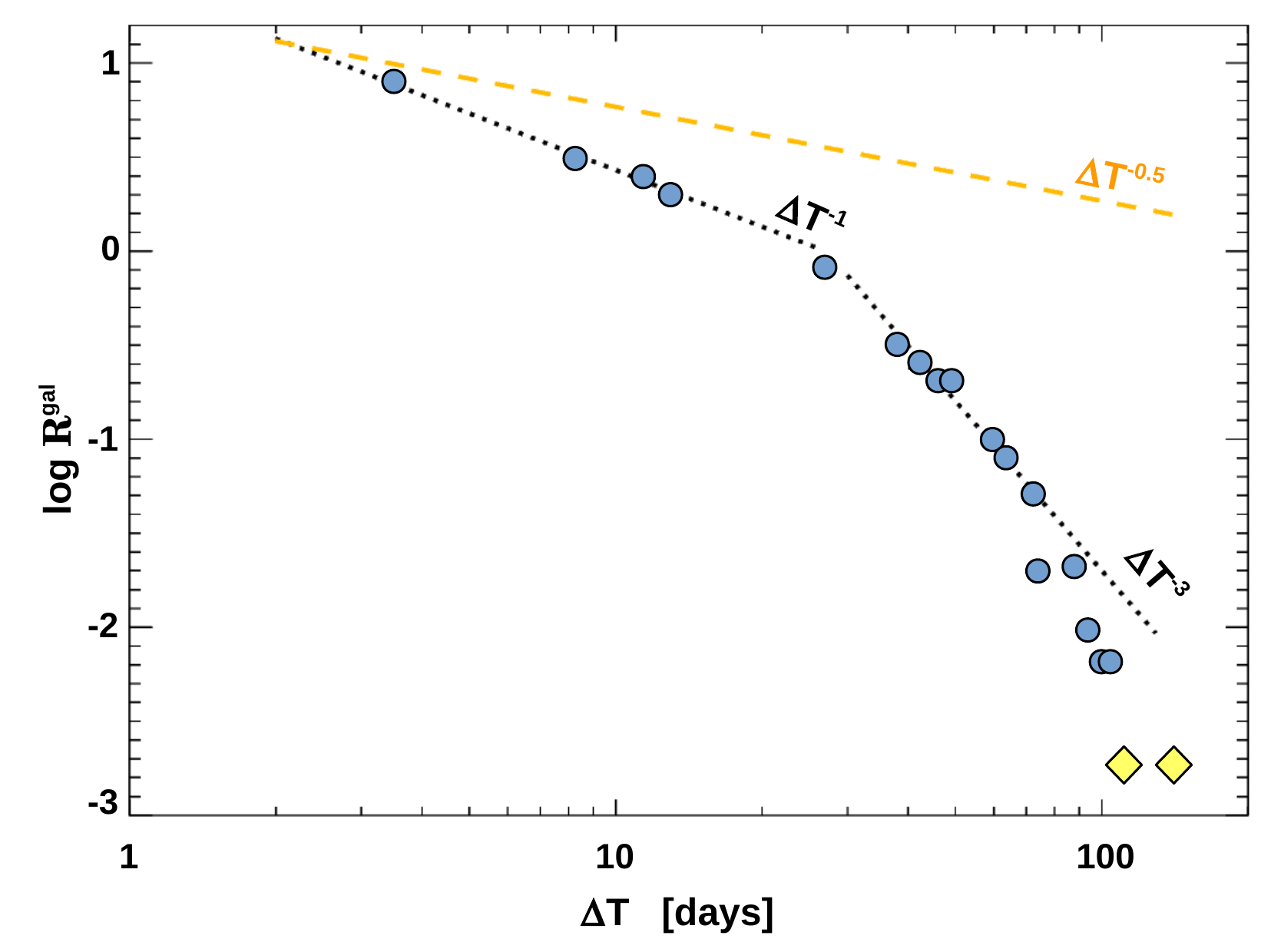} 
\caption{\label{fig2} Scaling of the LIGO-Virgo prior ratio on time delays, $\mathcal{R}^{\rm gal}$, with the time separation between GW events. The circles show the values tabulated in table 3 in \cite{LIGO2021}. The two diamond-shape yellow symbols at $\Delta T>100$ days have no values in table 3 of  \cite{LIGO2021}, and for presentation purposes we assign them a value of $log_{10}\mathcal{R}^{\rm gal} = -2.75$. 
The dotted lines and dashed line show the scaling $\mathcal{R}^{\rm gal} \propto \Delta T^{\alpha}$ for three values of $\alpha=-1,-3,-0.5$, with $\alpha=-1$ reproducing the LIGO-Virgo prior for time delays sorter than $\approx 1$ month, while $\alpha=-3$ better describes the prior for time delays longer than $\approx 1$ month. Overall, the LIGO-Virgo prior can be described with an exponent $\alpha=-2$. 
This should be compared with the value $\alpha=-0.5$, that reproduces well the observed time delay distribution for lensed QSOs as shown in Figure \ref{fig1}.
}
\end{figure}

The best clue about the adopted form for $\mathcal{R}^{\rm gal}$ is found in Table 3 in  \cite{LIGO2021}. In this table, surprisingly the pairs of events GW190421\_213856–GW190910\_112807 and GW190424\_180648–GW190910\_112807, having differences in arrival time of $\approx 4.7$ months  are given directly a probability of zero based on their time delay prior. An empirical prior based on the observed time delay in QSO would have not penalized these two pairs so heavily, compared with other events with shorter time delays (see Figure \ref{fig1}).  To do a more quantitative assessment, we use the values listed in Table 3 in \cite{LIGO2021}, to see how $\mathcal{R}^{\rm gal}$ scales with $\Delta T$. The result is shown in in \ref{fig2}. At time delays shorter than $\approx 1$ month  we find $\mathcal{R}^{\rm gal} \propto 1/\Delta T$, while for longer time delays we find $\mathcal{R}^{\rm gal} \propto 1/\Delta T^3$. If one considers the entire time delay range, the scaling $\mathcal{R}^{\rm gal} \propto 1/\Delta T^2$ is more appropriate. The change in slope may be due to the effect of $p(\Delta t|\mathcal{H}_{U})$, which makes longer time delays more consistent with the unlensed hypothesis. However, if the prior $p(\Delta t|\mathcal{H}_{SL})$ is already penalizing pairs of events with longer time delays (contrary to what observations of QSOs are telling us), it is not surprising that LVC concludes that these pairs of events are very unlikely to be lensed. From \ref{fig2}, it is evident that if one adopts the observational prior derived from the observed time delay in lensed QSOs (dashed orange line), the probability of the longer time delays considered by LIGO-Virgo would increase considerably. In figure \ref{fig2} we show also the scaling $\Delta T^{-0.5}$, which we find is the correct one in order to reproduce the observed distribution of QSO time delay. Based on all the above, we conclude that a better description for the prior term $\mathcal{R}^{\rm gal}$ would be an uninformative, or flat prior on $\Delta T$. Substantially more events need to be observed in order to discriminate between the distribution of time intervals from pairs of lensed candidates, and regular unlensed pairs.

%
% Made by LIGO/Plot_TimeDelay.pro in 2015 Toshiba computer

\section{Joint parameter estimation analysis}
%%%%%%%%%%%%%%%%%%%%%%%%%%%%%%%%%%%%%%%%%%%%%%%%
A more detailed analysis is presented in section 5.2 of \cite{LIGO2021}. This is a more computationally intensive analysis based on the 19 candidate pairs discussed in the previous section. In this case, the detector strain data is used instead of the posteriors to asses the similarity between GWs and the consistency of the 19 candidate pairs with the hypothesis that they are multiply imaged GWs. Two different analysis are presented. In the first one (\small{LALInference}), no prior assumption is adopted about the population of BBHs, nor lenses, and it takes into account the phase difference predicted by lensing \citep{Dai2017}. In the second analysis (\small{HANABI}), prior information is used about the population of BBHs and lenses. In particluar, for the BBH population, they adopt the maximum rate given by model A in their appendix A. This model, at the relevant redshift of z=1-2 predicts a rate that is approximately two orders of magnitude below the rate needed
to produce lensing events consistent with the observed rate of GWs with component masses in excess of 20 M$_{\odot}$. The Bayesian evidence presented in section 5.2 for the \small{HANABI} case is then biased by this low prior, and naturally the lensing hypothesis is disfavoured if one adopts a prior that intrinsically negates the possibility of lensing. A better and less biased result is obtained by their \small{LALInference} results, which avoids making assumptions about the pre-existence of lensing, and judges candidates based on the similarity of their strains and sky localization.

\section{Reinterpretating the massive LIGO-Virgo events as strongly lensed. The BDS model}\label{sec_BDS}
%%%%%%%%%%%%%%%%%%%%%%%%%%%%%%%%%%%%%%%%%%%%%%%%%%%%%%%%%%%%%%%%%%%%%%%%%%%%%%%%%%%%%%%%%%%%%%%%%%%%%%%%%%%
The LIGO-Virgo collaboration relies for its conclusions on the prior ratio  $\mathcal{R}^{\rm gal}$ (section 5.1) and  \small{HANABI} Bayes factor (section 5.2), but as argued above, serious consideration of lensing is effectively precluded by the choice of an idealised model prior for lensing time delays that is inconsistent with the measured QSO time delays and additionally, by reliance on a prior assumption that lensing generates only a minority of BBH events.  Instead, we have argued that one should not adopt those priors but instead rely on the Bayes factor $\mathcal{B}^{\rm overlap}$ in section 5.1 of \citep{LIGO2021}, or the more precise coherence ratio from the \small{LALInference} pipeline in section 5.2 of \citep{LIGO2021}, for the comparison of pairwise waveforms and sky positions. This leads us to conclude that 11 out of the original 19 LVT candidate lensed pairs are in fact highly consistent with the lensing hypothesis, not only in terms of waveforms and sky location but also in terms of the known distribution of QSO time delays. We take this conclusion as firm evidence in favour of the lensing hypothesis, that significant numbers of lensed GW events are present in the published O3a catalog. Furthermore, there is at least one obvious triple set of coincident pairs worth highlighting within the 19 pairs identified by \citep{LIGO2021}, namely, GW190731,GW190803 \& GW190910. This is interesting because the larger the multiplicity of detected events, the lower is the corresponding BBH coalescence rate required by lensing to match the data.
\\

In earlier work we proposed an alternative model for the mass function and rate of mergers that predicts a significant number of lensed events in LIGO-Virgo observations. We refer to this model as the BDS model \citep{Broadhurst2018,Broadhurst2019,Broadhurst2020a,Broadhurst2020b}. Here we briefly review the main properties of this model. 
In order for strong lensing of GWs to be observed at current detector sensitivities, the rate of mergers at high redshift must be sufficiently large so it can compensate for the small optical depth of lensing. Accordingly, the rate at low redshift is reduced  relative to LVT estimates, as we conclude most events to date are lensed at much higher redshift.

The BDS model relies on the known mass distribution of stellar black holes established before the LVT era, and the universal magnification relation for lensing caustics to derive the distribution of lensed events. The only important free parameter is the evolution of the mass function, as lensing preferentially accesses high redshift events, and hence this early rate is what we primarily constrain using the current data.\\

Inspired by the results from X-ray binaries in the Milky Way\citep{Corral-Santana2016}, we parameterize the mass function as a log-normal 
\begin{equation}
    \frac{dN}{dM} = \frac{1}{M\,\sigma_M}exp\left( \frac{-(log_{10}(M)- \mu_M)^2}{2\, \sigma_M^2} \right)
    \label{eq_massfunct}
\end{equation}
with a mean $\mu_M=log_{10}(8)$ and dispersion $\sigma _M = 0.27$ \Msun. Since at high redshift massive stars are more prevalent, we allow for a mild evolution in the mean mass with redshift and assume the mean above $z=1$ is $\mu_m=log_{10}(13)$, and keep the dispersion to $\sigma_M = 0.27$ \Msun. For the rate of mergers as a function of redshift we assume a model with a low rate at low redshift and a high rate at high redshift. In particular, the model is given by the expression:

\begin{equation}
\label{eq_rate}
R(z)=\left\{
\begin{aligned}
 A_1exp\left( \frac{T(z)-T(z_{max})}{T_H} \right) \quad {\rm IF} \  z<z_{max}\\
\frac{A_2(1+z)^{2.7}}{1+(\frac{1+z}{2.9})^{5.6}}\quad {\rm IF}\  z>z_{max}
\end{aligned}
\right.
\end{equation}
Where $T(z)$ is the time in Gyr at redshift $z$, the amplitude $A_1$ corresponds to the merger rate at redshift $z_{max}$, the half-life time $T_H=1.25$ Gyr, and the normalization $A_2$ is adjusted in such a way that at $z=z_{max}$ the rate given by the expression for $z>z_{max}$ is equal to $A_1$, that is $A_2=A_1\, [1+((1+z_{max})/2.9)^{5.6}]/(1+z_{max})^{2.7}$. The merger rate evolves rapidly (exponentially) between redshifts $z=0$ and $z=z_{max}$, where it peaks. Above redshift $z=z_{max}$, the rate is assumed to decline more slowly following the cosmic star formation rate \citep{Madau2014}. Since the expression above $z=z_{max}$ peaks at $z\approx 1.8$, we adopt $z_{max}=1.8$. The redshift $z_{max}$ can be modified and still reproduce the observed rate of events by LIGO. Values of the rate at z=1.8 of $A_1\approx 3\,.\,10^4$ yr$^{-1}$Gpc$^{-3}$ are needed in order to reproduce the observed distribution of BH masses. A comparison of a model similar to Eq.~\ref{eq_rate} with a model that traces the star formation rate can be found in Figure 3 of \cite{Diego2020}. 
A larger number of GW detections will allow in the future to constrain this amplitude better, together with $T_H$ and $z_{max}$.  

\begin{figure}[ht]
\includegraphics[width=9cm]{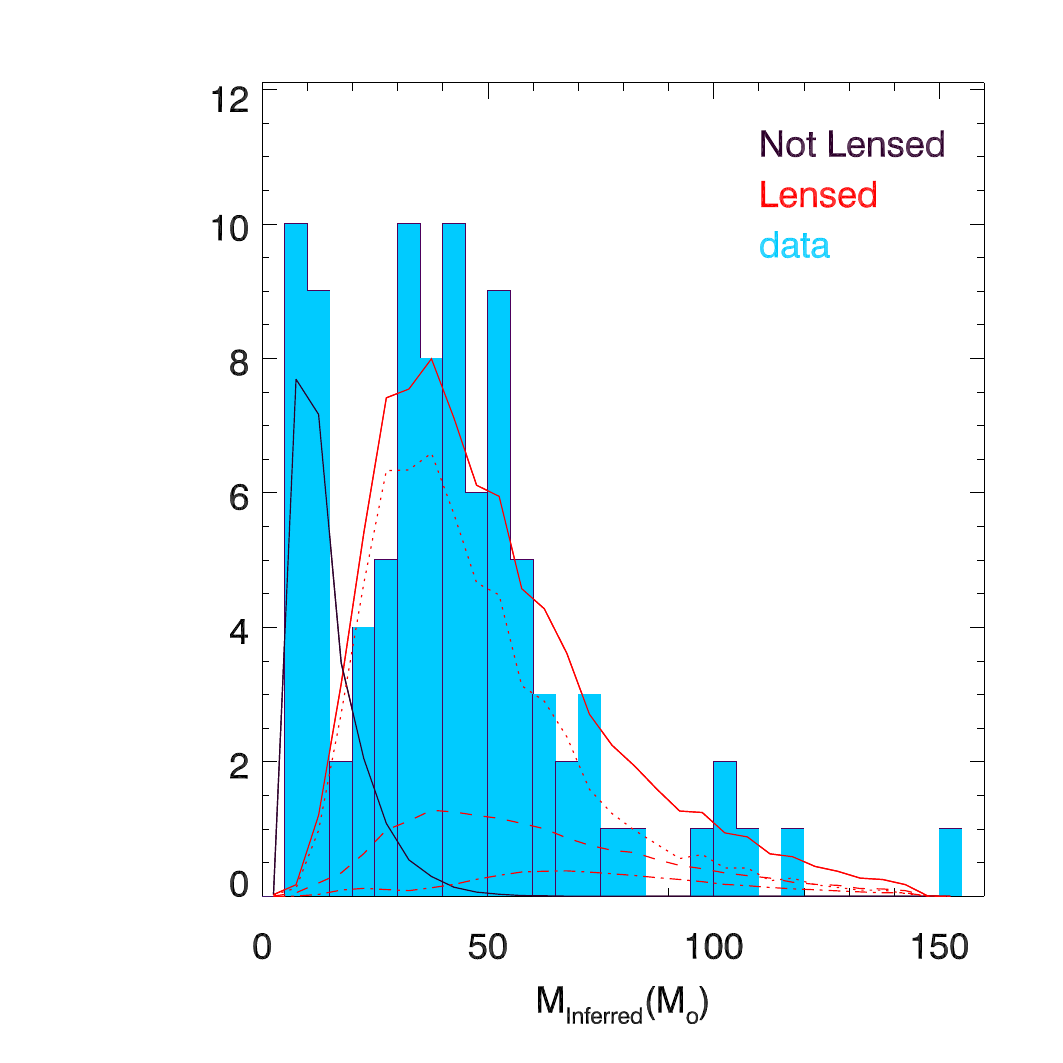} 
\caption{\label{fig3} Histogram of inferred black hole masses published by the LIGO/Virgo collaboration vs expectations from the BDS model. The blue bars show the histogram of published masses. The solid dark blue curve shows the expected number of not-lensed events from the BDS model in one year of LIGO/Virgo observations at O3 sensitivity. In this case the inferred mass is unbiased, and the distribution of masses follows equation \ref{eq_massfunct}. The solid red line shows the corresponding number of events that are being lensed from the same model, and for the same sensitivity and observation period. In this case, the inferred mass is biased due to lensing being ignored (the inferred redshift is smaller than the true one and this bias is compensated by a larger inferred mass). The dotted, dashed, and dot-dashed lines correspond to the number of events from gen11, gen12, and gen22 respectively. For clarity these curves are not shown for the not-lensed case. 
}
\end{figure}

The final ingredient in the model is given by the optical depth of lensing, which in its most general form is a function that depends on the redshift of the source and the magnification factor, $\tau(z_s,\mu)$. For $\tau(z_s,\mu)$ we adopt the results from \cite{Diego2020}, which find for instance an optical depth of $\tau \approx 1.5\, 10^{-4}$ for events with magnification $\mu>5$ at $z_s=2$ (see section 3 in that reference).   
For this mass function, merger rate, and optical depth we find that the massive events observed by LIGO ($M_{BH}>20$ \Msun) can be reinterpreted as strongly lensed events from binaries originating at $z>1$, as shown in our earlier work. For instance, the inferred mass function of BH from LIGO-Virgo is clearly bi-modal as shown in figure \ref{fig3}. This bimodality is naturally explained under the lensing interpretation, with the first narrow peak at low mass corresponding to the local events that are not being strongly lensed, and the second and broader peak at larger masses corresponding to the more distant strongly lensed events. The predictions from our model are shown as solid lines. The blue line represents the events that are not lensed, and the red curve represents the events that are strongly lensed. We show three additional curves for events with two generation 1 BHs (or gen11 shown as a doted line), mergers with a generation 1 and a generation 2 BHs\footnote{A generation 2 BH is the result of the merger of two generation 1 BHs} (or gen12 shown as a dotted line) or mergers with two generation 2 BHs  (or gen22, shown as a dot-dashed line). For this plot we have assumed generation 2 BHs are 9 times less abundant than generation 1. A much smaller fraction of generation 2 BHs results in similar results (see the dotted line), but the presence of generation 2 BHs improves the fit for the most massive events above $\approx 80$ \Msun.  Our population of generation 2 BHs is obtained by randomly extracting BH masses from the mass function in equation \ref{eq_massfunct}. 
%The probability of an event genij is given by the product of their number density, i.e $\rho_i\, .\, \rho_j$, 

%
% Made by LIGO/plot_Histogram in 2018 Dell computer

We can conclude then that if LIGO is observing a significant number of gravitationally lensed events, the rate of mergers at $z>1$ must be in the range of  $\approx 10^4$--$10^5$ yr$^{-1}$Gpc$^{-3}$. Lensing would offer an alternative interpretation for the high masses found by the LIGO-Viorgo collaboration.

So far we have shown in our earlier work how the simple BDS model naturally explains the basic observed features of the LVT detection, namely; \\
$\bullet$ The consistency between the black hole mass function observed in our Galaxy and the one inferred from gravitational wave observations \citep[lensed GWs from distant binaries with masses similar to the BHs found in our Galaxy would be incorrectly interpreted as more massive black holes;][]{Broadhurst2018}, \\
$\bullet$ the bimodal mass function distribution of BHs \citep[the low-mass peak corresponds to the local non-lensed GWs, and the high mass peak to the distant lensed GWs;][]{Broadhurst2019},\\
$\bullet$ the tight correlation between the two mass components of the binary, $m_1$ and $m_2$ \citep[both $m_1$ and $m_2$ are drawn from a narrow log-normal mass function peaking at around 8 M$_{\odot}$, where the difference $m_1-m_2$ is naturally small;][]{Broadhurst2020a}, \\
$\bullet$ the existence of black holes in the mass gap \citep[the neutron star in a NSBH binary at redshift $1 < z < 2$ which is being lensed, would be interpreted as having a mass between 3 and 5 M$_{\odot}$, or a BH with mass M $=15$ M$_{\odot}$ at $z>3$, would be incorrectly interpreted as a M $> 50$ M$_{\odot}$ at $z<1$;][]{Broadhurst2020b}. \\

In this paper we argue that despite the claims made recently by the LIGO-Virgo team, the recent work presented in \cite{LIGO2021} strengthens the case for lensing. Consequently,  an additional bullet should be added to the list above; \\
$\bullet$ The consistency we find here between the distribution of time delays known for lensed QSOs. And the time differences between pairs of BBH events that have been classified by the LVT as overlapping on the sky with compatible waveforms, as required by lensing (time delays between observed GWs with compatible sky localization are consistent with expectations, providing additional evidence for lensing; this work). \\

\section{Conclusions}\label{se_concl}
%%%%%%%%%%%%%%%%%%%%%%%%%%%%%%%%%%%%%%%%%%
We have discussed how the prior ratio $\mathcal{R}^{\rm gal}$ may be ill-defined, and likely biases the conclusions derived in \cite{LIGO2021}. A prior drawn from actual observations of time delays in QSOs, results in a very different picture, with pairs of events being separated by up to months being quite consistent with the lensing hypothesis, instead of being heavily penalized by the rapidly decaying SIS model prior of the LVT analysis. The high degree of parameter consistency found by the LIGO-Virgo collaboration for 11 pairs of BBH events is suggestive that lensing of GWs may be taking place at a much higher rate than previously realised by the LVT, but as predicted by the BDS model. If confirmed, and in light of this, it would imply that models with large merger rates at $z>1$, like in the BDS model, are necessary in order to explain the coincidences both in space parameter and sky location found by the LIGO-Virgo collaboration (and quantified by  $\mathcal{B}^{\rm overlap}$ and $C^L_U$). If the high rates at $z>1$ predicted by the BDS model are correct, the evidence for lensing can only grow stronger as more data is analyzed, with increased coincidences in strain parameters, and in particular of course with improved sky location using multiple detectors. If the BDS model is correct, it predicts that the stochastic background of GWs will be detected in the near future, and will differ in shape from the predictions made by more standard models \citep[see][]{Mukherjee2021}, that assume a much higher rate at low redshift and lower rate at high redshift, than the BDS model.   \\

The LIGO-Virgo collaboration has already confirmed good consistency between strains and sky location for 11 pairs of GW events in the released first half of the O3 events (O3a), but a more definitive outcome may be provided by the longer duration of the full O3 set of events which should allow the distinctive empirically based time delay distribution to be recognised more clearly with an expected peak at $\simeq 100$ days based on QSO evidence. Finally, in order to properly rule out the lensing hypothesis, after incorporating priors on the BBH population and lenses,  models that do predict observable rates of lensed events (like the BDS model) need to be properly considered.  This model may eventually be proven wrong by future data, but current observations are still consistent with it, that predict that lensing of binaries with components in the $\approx 7$--$15$ M$_{\odot}$ at $z>1$ are responsible for the bimodal mass function, and also the tight correlation between $m_1$--$m_2$, mass-gap events, and the similarity in strains and sky locations between a significant number of GW events.

\begin{acknowledgments}
%%%%%%%%%%%%%%%%%%%%%%%%%
We thank James Chan for useful discussions and for providing the QSO time delay catalogue, and the anonymous referee for constructive feedback. J.M.D. acknowledges the support of project PGC2018-101814-B-100 (MCIU/AEI/MINECO/FEDER, UE) Ministerio de Ciencia, Investigaci\'on y Universidades.  This project was funded by the Agencia Estatal de Investigaci\'on, Unidad de Excelencia Mar\'ia de Maeztu, ref. MDM-2017-0765. 

\end{acknowledgments}

% The \nocite command causes all entries in a bibliography to be printed out
% whether or not they are actually referenced in the text. This is appropriate
% for the sample file to show the different styles of references, but authors
% most likely will not want to use it.
\nocite{*}

\bibliography{MyBib}% Produces the bibliography via BibTeX.

\end{document}